\documentclass[11pt,twoside]{article}
\usepackage{asp2006}
\usepackage{graphicx}
\begin{document}

\title{Observations of the Joint Action of the Hanle and Zeeman Effects in the D$_2$ Line of Ba~{\sc{ii}}}

\author{R.~Ramelli,$^1$ M.~Bianda,$^{1,4}$ 
J.~Trujillo Bueno,$^{2,5}$ L.~Belluzzi,$^3$ and 
E.~Landi Degl'Innocenti$\,^3$}

\affil{$^1$Istituto Ricerche Solari Locarno, CH-6605 Locarno, Switzerland}    

\affil{$^2$Instituto de Astrof\'{\i}sica de Canarias, 38205, La Laguna, Tenerife, Spain}    

\affil{$^3$Universit\`a degli Studi di Firenze, Dipartimento di Astronomia e
Scienza dello Spazio, Largo Enrico Fermi 2, I-50125 Firenze, Italy}    

\affil{$^4$Institute of Astronomy, ETH Zurich, CH-8092 Zurich}

\affil{$^5$Consejo Superior de Investigaciones Cient\'\i ficas, Spain}

\begin{abstract} 
We show a selection of high-sensitivity spectropolarimetric
observations obtained over the last few years in the Ba
{\sc{ii}} D$_2$-line with the
Z\"urich Imaging Polarimeter (ZIMPOL) attached to the Gregory Coud\'e
Telescope of IRSOL. The measurements were collected close to the solar
limb, in several regions with varying degree of magnetic
activity. The Stokes profiles we have observed show clear signatures of the joint action
of the Hanle and Zeeman effects, in very good qualitative agreement
with the theoretical expectations. Polarimetric measurements of this
line show to be very well suited for magnetic field diagnostics of the lower solar chromosphere, 
from regions with field intensities as low as 1 gauss to strongly magnetized ones having kG field strengths.
\end{abstract}

\section{Introduction}   
In a recent paper, \citet*{ramelli_2007ApJ...666..588B} concluded that the modeling
of spectropolarimetric observations in the Ba {\sc{ii}} D$_2$-line at
4554 \AA\ 
could be a very useful diagnostic tool to obtain empirical information
on the spatial fluctuations of the magnetic field vector in the lower
solar chromosphere. This line is indeed particularly interesting
because its emergent linear polarization has contributions from
different isotopes that 
have a different
behavior in the presence of a magnetic field.

With the observations reported in the present paper we 
aim at experimentally verifying the theoretical model predictions 
on the magnetic sensitivity of the Ba {\sc ii} D$_2$ line.

\section{Observations}
A set of 28 spectropolarimetric measurements were 
obtained at several regions close to the solar limb
during 9 different days from June 2005 to March 2007 with the 45 cm
Gregory-Coud\'e Telescope (GCT) at  the Istituto Ricerche Solari Locarno (IRSOL).
The observations were obtained at different latitudes, from  the equator to
the solar poles, and in both quiet and active regions.
A high polarimetric sensitivity free from seeing induced effects could be
achieved thanks to the ZIMPOL polarimeter \citep{ramelli_hp,ramelli_gandorfer04}. 
The solar image was rotated with a Dove prism set after the polarization
analyzer in order to keep the limb parallel to the spectrograph slit. 
A limb tracker was used to keep a constant distance from the limb:
typically 10 arcsec. Calibration observations such as
polarimetric efficiency measurements, dark current, flat field, and
instrumental polarization measurements were taken regularly before
and/or after the observations.  The instrumental polarization was carefully corrected for applying the procedure
described in \citet{ramelli_2006ASPC..358..448R}.

\section{Results}   
The observations exhibit a very good qualitative agreement with the basic
theoretical predictions of \citet{ramelli_2007ApJ...666..588B}.
We report here a few significative examples. 

\begin{figure*}[!b]
\begin{center}
\includegraphics[angle=0,width=.48\linewidth,height=6.8cm]{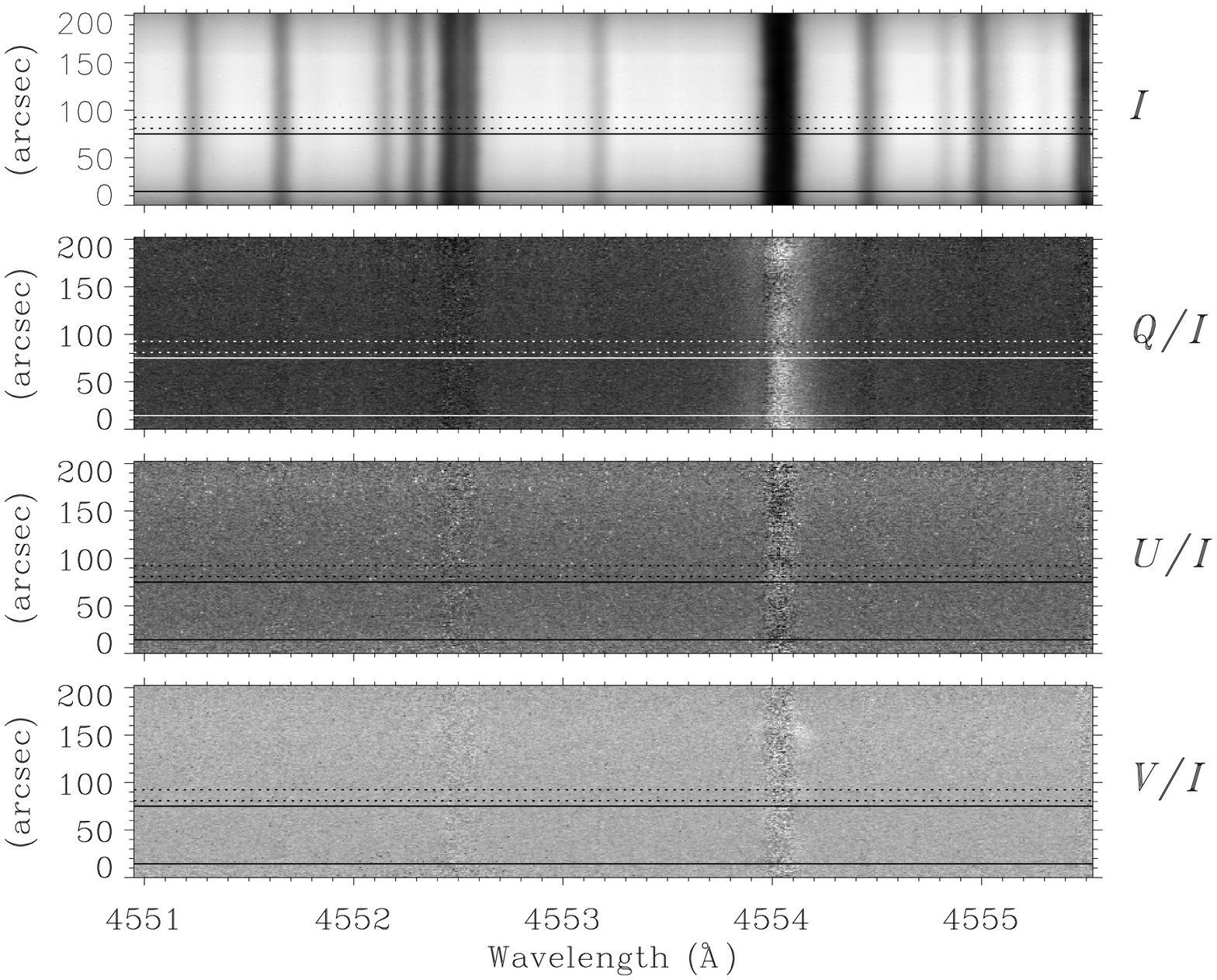}
\hfill
\includegraphics[angle=0,width=.48\linewidth,height=6.8cm]{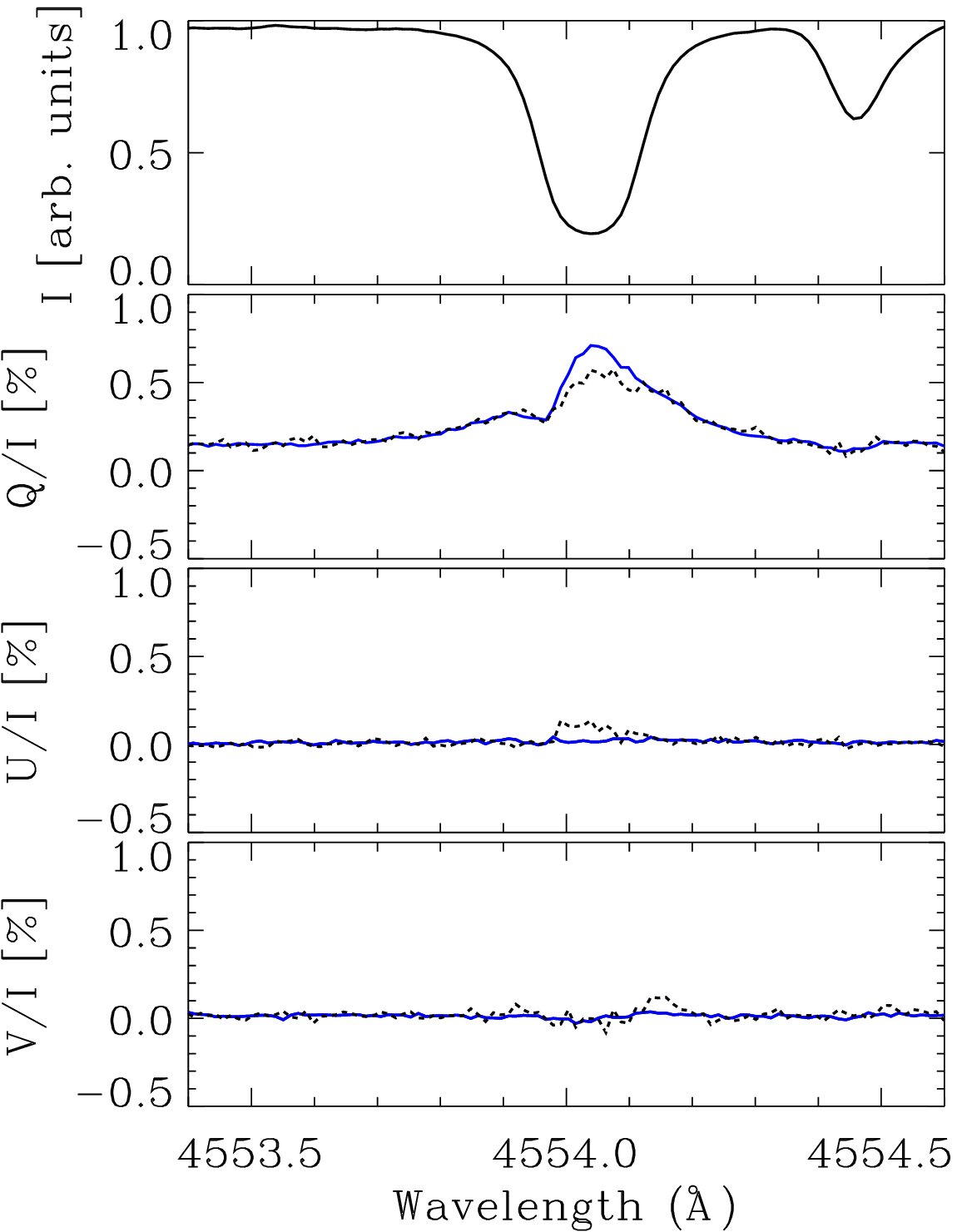}
\caption{Full-Stokes observations taken on
August 15, 2006 close to the heliographic North Pole with the 
spectrograph slit at about 10 arcsec from the limb. Two spatial
regions are selected by the solid and dotted horizontal lines on the
Stokes images reported in the left panel. The corresponding Stokes
profiles around the Ba~{\sc ii}~D$_2$ line are shown on the right panel by the solid and dotted
lines. The latter exhibit the typical behavior expected for a 
weak horizontal magnetic field  with a significant 
longitudinal component (${\sim}$ 1 gauss). 
The exposure time was 800 s. 
}
\label{ramelli_fig1}
\end{center}
\end{figure*}

The measurement reported in Fig.~\ref{ramelli_fig1} was obtained 
 close to
the heliographic North Pole. Two spatial regions along the
spectrograph slit are selected and the corresponding profiles shown on
the right panel. In the region corresponding to the dotted
lines,  there is a small decrease of the
$Q/I$ signal in the line core while the line wings remain
unaffected. In the same region a small feature appears in $U/I$ in the
line core. The $V/I$ profile shows a small antisymmetric signal.
This is the typical
behavior expected for  a horizontal magnetic field 
with a longitudinal component of
the order of 1 gauss (compare with fig. 10 of Belluzzi et al. 2007).
 
\begin{figure*}[!b]
\begin{center}
\includegraphics[angle=0,height=7cm]{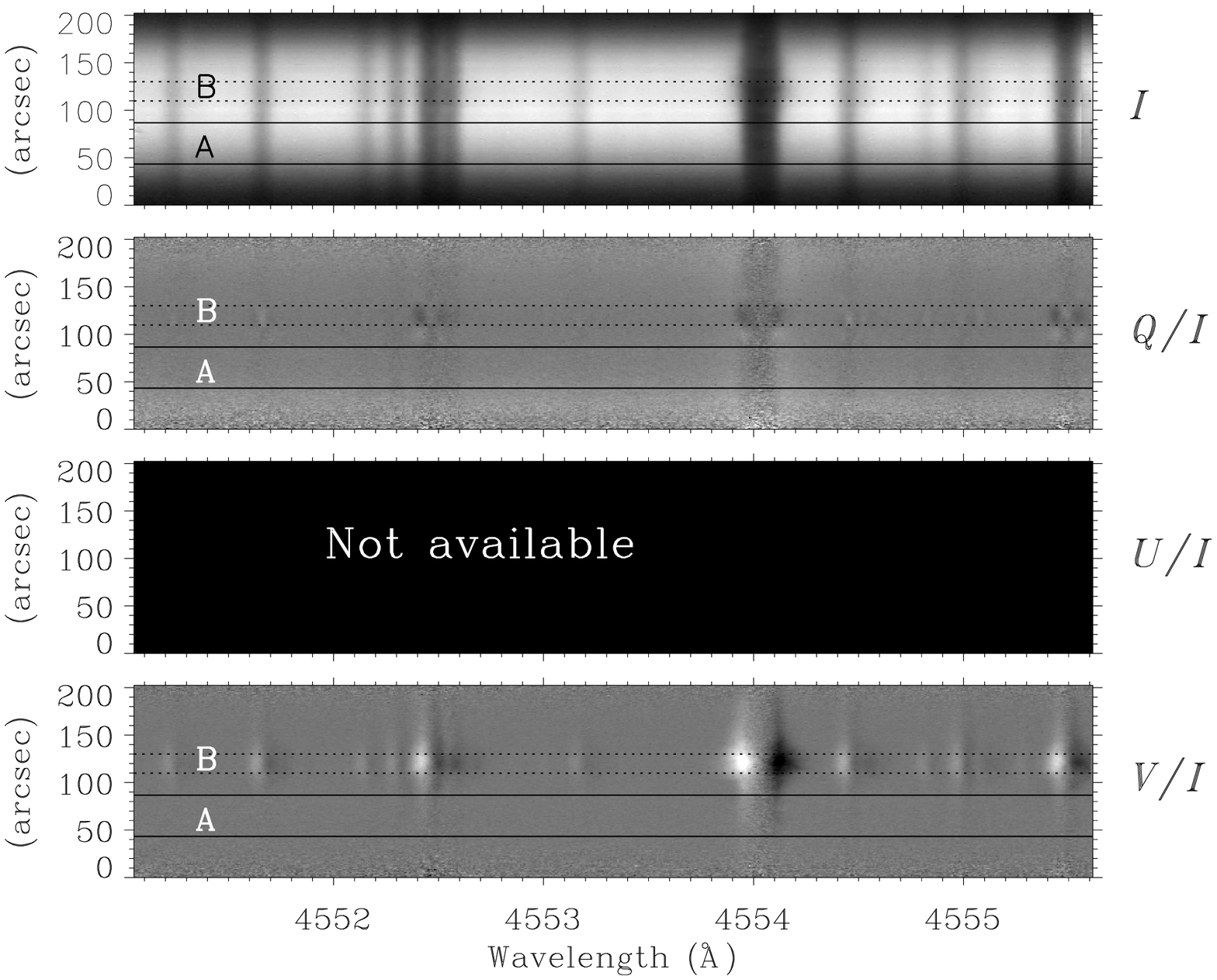}
\hspace{1.2cm}
\includegraphics[angle=0,height=6.2cm]{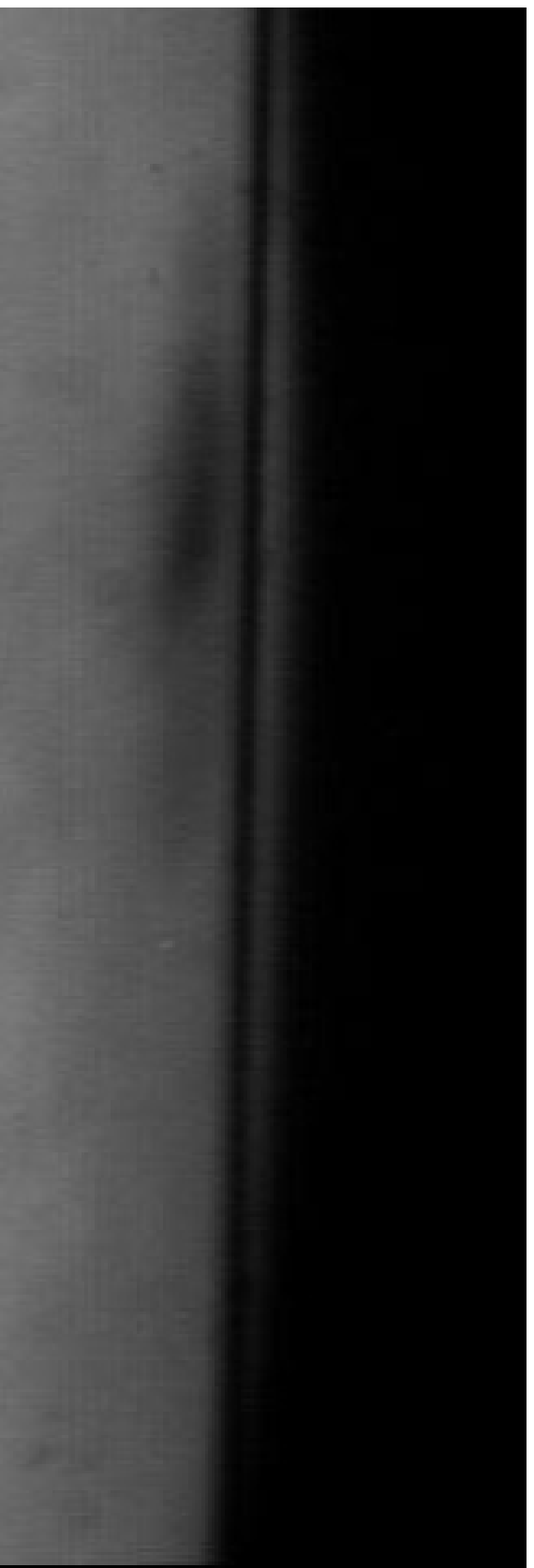}
\caption{Observations taken on August 21, 2006 at about 3 arcsec
from the 
 limb, very close to a sunspot. The left panel shows 
the observed $I$, $Q/I$ and $V/I$ images. Two spatial regions $A$
and $B$ are selected. 
 The corresponding profiles are shown in Fig.~\ref{ramelli_fig3}.
In Region $A$ there is a 
 magnetic field of intermediate strength. Region
 $B$ is selected where the magnetic field strength is at the maximum,
 close to the sunspot.
 A cropped low resolution
 slit-jaw image taken during the measurement is shown on the right. 
 The total exposure time was 1200 s.}
\label{ramelli_fig2}
\end{center}
\end{figure*}

\begin{figure*}[!ht]
\begin{center}
\includegraphics[angle=0,width=.65\linewidth]{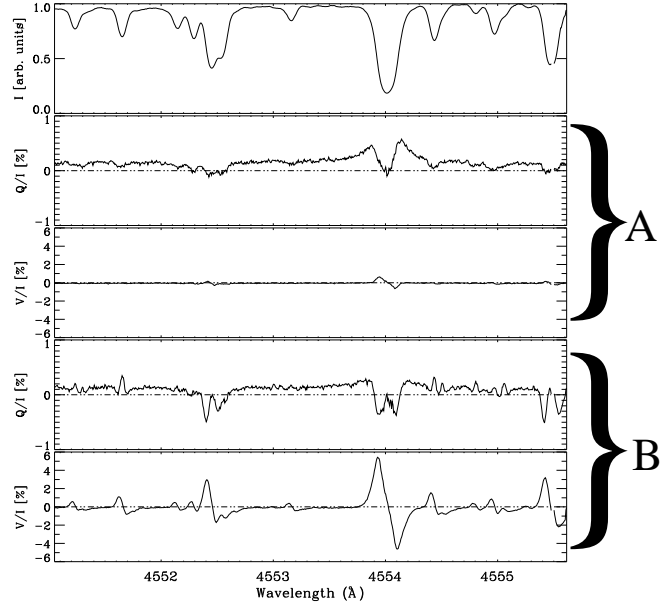}
\caption{$I$ profile, $Q/I$ profile and $V/I$ profile of regions $A$ and
$B$ selected in Fig.~\ref{ramelli_fig2}}
\label{ramelli_fig3}
\end{center}
\end{figure*}

In the left panel of Fig.~\ref{ramelli_fig2} 
we show a measurement of
Stokes $I$, $Q/I$ and $V/I$ taken near the west limb  (AR 10904)
 very close to
a sunspot (see the slit jaw image on the right panel of Fig.
\ref{ramelli_fig2}). The 
corresponding profiles obtained in two selected spatial regions $A$
and $B$ along the slit are displayed in Fig.~\ref{ramelli_fig3}.  Region $A$
corresponds to a 
region with an intermediate field strength. Here
we see again in the $Q/I$ profile a decrease of polarization in the
line core while the line wings stay almost unaffected: the typical
behavior expected for a horizontal magnetic field with a strength smaller than
100 Gauss (compare with the left panels of fig. 10 of Belluzzi et al. 2007). 
Region $B$ is selected just above the sunspot where the Stokes $V/I$
Zeeman signals reach the maximum amplitude. In this case one observes 
a strong decrease of $Q/I$ in the line wings due to the transverse
Zeeman effect. This is well expected for a magnetic field whose
horizontal component perpendicular to the line of sight is larger than
100 gauss (see the right panels of figs. 7 and 9 in Belluzzi et al. 2007).

\begin{figure*}[!ht]
\begin{center}
\includegraphics[angle=0,width=.8\linewidth]{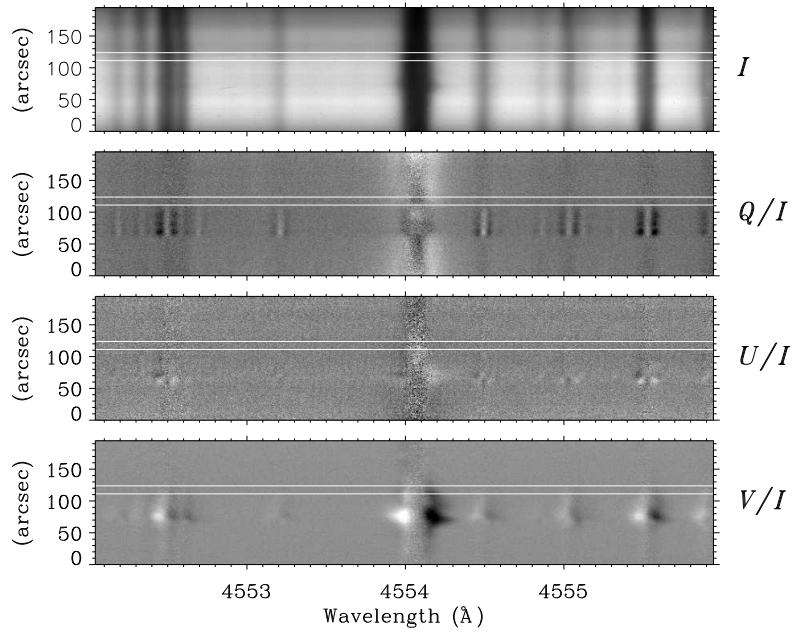}
\caption{Full-Stokes observations taken on March 13, 2007 in an
active region with the spectrograph slit set at about 8 arcsec from the
limb. The exposure time was 4400 s. Profiles obtained in different
portions of the slit are reported in Fig.~\ref{ramelli_fig5}. The profiles
obtained in the interval between the two horizontal lines present in
each Stokes image are reported in Fig.~\ref{ramelli_fig6}}
\label{ramelli_fig4}
\end{center}
\end{figure*}

\begin{figure*}[!ht]
\begin{center}
\includegraphics[angle=0,width=.49\linewidth]{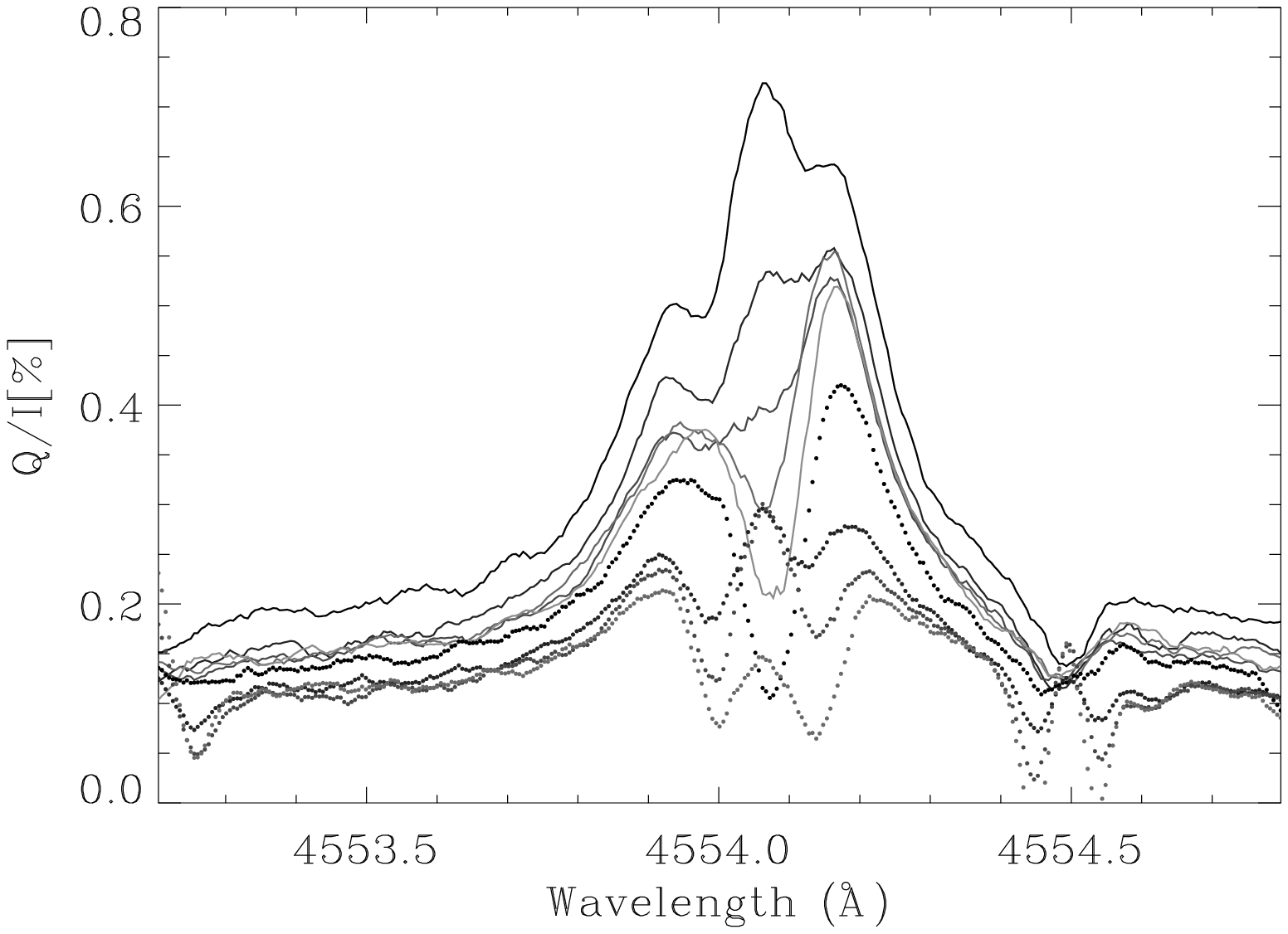}
\hfill
\includegraphics[angle=0,width=.49\linewidth]{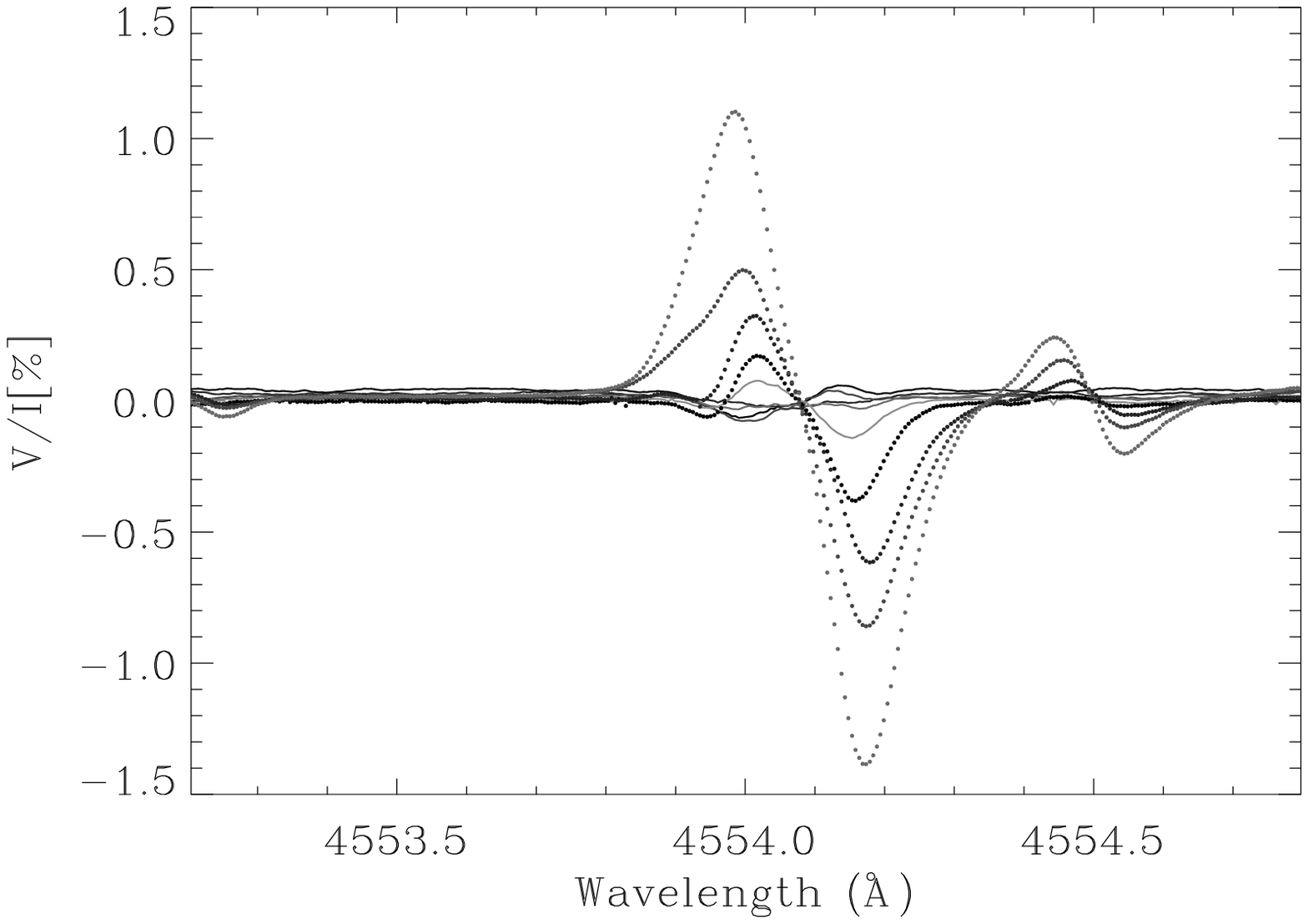}
\caption{Stokes $Q/I$ and $V/I$ profiles of different contiguous spatial regions
of about 15 arcseconds each, selected from the top of the image shown in Fig.
\ref{ramelli_fig4} down to the region with the strongest Stokes $V/I$
signal. In the transition from the top of the image to the latter
position, the Stokes $Q/I$ signal reduces gradually its amplitude while
the antisymmetric Zeeman pattern in $V/I$ increases
its amplitude.}
\label{ramelli_fig5}
\end{center}
\end{figure*}

Finally we show the results of a long exposure measurement (4400 s) taken 
at about 8 arcsec from the limb, just above the active region AR 10946
on March 13, 2007 (Fig.~\ref{ramelli_fig4}). Due to the high photon statistics
it is possible to obtain several good quality profiles, with low level of
noise, integrating over small portions of the spectrograph slit. 
Starting from the top of the image we select several contiguous regions of 
about 15 arcseconds (10 pixels) down to the region with the highest Stokes
$V/I$ signal. The resulting $Q/I$ and $V/I$ profiles are shown in
Fig.~\ref{ramelli_fig5}. This illustrates quite well a transition from a 
weakly magnetized region to a substantially more active one.
In $V/I$
there is a gradual increase of the amplitude of the antisymmetric Zeeman
pattern. In $Q/I$ one sees first a clear decrease of polarization in the
line core and then, where the field becomes stronger, also a decrease in the 
line wings. One notes also an overall decrease of $Q/I$ on the whole
spectral range. The larger overall value of $Q/I$ in the top of the
image is due to the fact that we are closer to the limb than in the
center of the image because of the limb curvature.


\begin{figure*}[!ht]
\begin{center}
\includegraphics[angle=0,width=.61\linewidth]{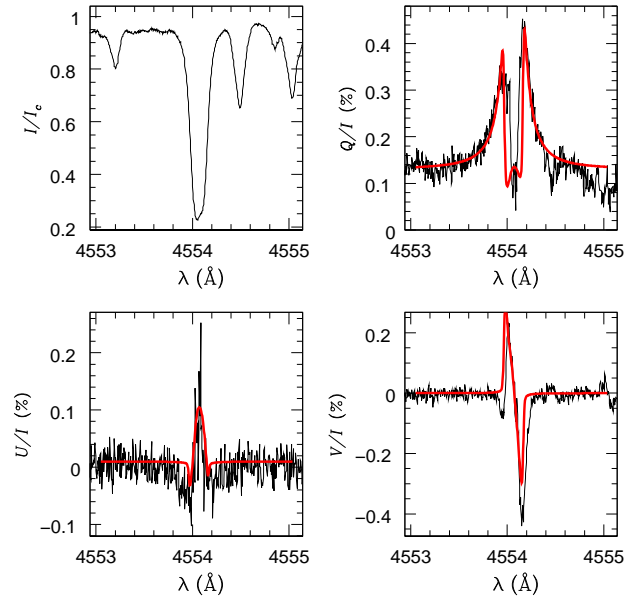}
\caption{An example showing the profiles obtained in the 15
 arcseconds region included between the line pairs shown in Fig.
 \ref{ramelli_fig4}. The files are fitted with theoretical profiles predicted
 by the model of \citet{ramelli_2007ApJ...666..588B} corresponding to a
 magnetic field of 28 Gauss with an inclination $\Theta_B=63^\circ$
 with respect to the local vertical
 and an azimuth $\chi_B=98^\circ$ (for the angle definitions see the
 corresponding paper).
}
\label{ramelli_fig6}
\end{center}
\end{figure*}

The profiles obtained in this observation were fitted with the
theoretical profiles predicted by the optically-thin model of
\citet{ramelli_2007ApJ...666..588B}, which is however 
very detailed from the atomic physics viewpoint.
An example of the result of the fit is shown in Fig.~\ref{ramelli_fig6}. 
Obviously, the goodness of this fit does not imply at all the validity 
of the optically thin assumption, which was used on purpose by \citet{ramelli_2007ApJ...666..588B}
in order to focus on understanding the magnetic sensitivity of the emitted spectral line polarization.
Since the Ba {\sc ii} D$_2$ line is strong, any quantitative modeling of the present observations will have to
include the effects of radiative transfer in realistic atmospheric models.

\section{Conclusions}
The high-sensitivity spectropolarimetric observations 
reported in the present paper are in 
good qualitative agreement with some of the theoretical predictions of 
\citet{ramelli_2007ApJ...666..588B} on the magnetic sensitivity of the Ba {\sc ii} D$_2$ line.
This reinforces their conclusion that 
spectropolarimetric measurements of this line provide a 
very promising mean for magnetic field diagnostics in the solar
photosphere and chromosphere for a wide range of field strengths.
Particularly interesting results may be expected with two-dimensional
spatial polarimetric imaging techniques using Fabry-Perot interference
filters.

\acknowledgements
Financial support by the canton of Ticino, the city of Locarno, the ETH Zurich,
SNF grant
200020-117821, and the Spanish Ministry of Science through project AYA2007-63881 is gratefully 
acknowledged.
 Thanks
are also due to the ETH ZIMPOL group for the technical support.

\end{document}